\documentclass[12pt]{article}
\usepackage{a4}
\newcommand{\beao}{\begin{eqnarray*}}
\newcommand{\eeao}{\end{eqnarray*}}
\newcommand{\be}{\begin{equation}}
\newcommand{\ee}{\end{equation}}
\newcommand{\bea}{\begin{eqnarray}}
\newcommand{\eea}{\end{eqnarray}}
\newcommand{\beq}{\begin{eqnarray}}
\newcommand{\eeq}{\end{eqnarray}}
\newcommand{\nn}{\nonumber}

\newcommand{\la}{\lambda}

\newcommand{\Ref}[1]{(\ref{#1})}

\usepackage{epsfig}
\begin{document}
\title{Spontaneous magnetization of the vacuum \\ and the strength of the magnetic field \\ in the hot Universe}
\author{E.~Elizalde$^{a}$\thanks{E-mail: elizalde@ieec.uab.es, elizalde@math.mit.edu}~
 and~V.~Skalozub$^{b}$\thanks{E-mail: skalozubv@daad-alumni.de}
 \\
  {\small\textit{ $^a$Institute for Space Science, ICE-CSIC and IEEC}}\\
 {\small\textit{ Campus UAB, 08193 Bellaterra, Barcelona, Spain}} \\
 {\small\textit{ $^b$Dnipropetrovsk National University, 49010 Dnipropetrovsk, Ukraine}} \\}
%
\maketitle
\begin{abstract}
Intergalactic magnetic fields are assumed to have been spontaneously
generated at the reheating stage of the early Universe, due to vacuum
polarization of non-Abelian gauge fields at high temperature. The fact that
the screening mass of this type of fields has zero value was discovered
recently. A procedure to estimate their field strengths, $B(T)$, at
different temperatures is here developed, and the value
$B(T_{ew}) \sim 10^{14} G$ at the electroweak phase transition
temperature, is derived by taking into consideration the present
value of the intergalactic magnetic field strength, $B_0 \sim 10^{- 15} G$,
coherent on the $\sim 1$ Mpc scale. As a particular case, the standard
model is considered and the field scale at high temperature is estimated
in this case. Model dependent properties of the phenomena under
investigation are briefly discussed, too.
\end{abstract}
\section{Introduction}
The recent experimental discovery of intergalactic magnetic fields
having a field strength of the order $B \sim 10^{-15}G$ is  one of
the most interesting events of modern cosmology \cite{Ando,Neronov}.
In Ref.~\cite{Ando2} a model-independent, 95\%
CL interval  $1 \times 10^{-17} G \leq B \leq 3 \times
10^{-14} G $  was determined, and the femtoGauss values are actual
field strengths in extragalactic space. This means, to start,
that magnetic fields are actually present everywhere in the Universe and
influence various processes. Secondly, this renders most likely
the cosmological origin of primordial magnetic fields. From the
theoretical viewpoint, this discovery  restricts in an
essential way the possible processes which may result in the creation
of fields in the hot Universe \cite{Neronov}. As a consequence, the search
for mechanisms of field generation has intensified. The most obvious
candidates are primordial fluctuations, but there
are other (for a review, see \cite{Grasso}-\cite{Kunze}). The challenge
is to produce  coherent magnetic fields on very long scales in an almost
empty intergalactic space. In this paper we will discuss some mechanism based
on non-Abelian magnetic fields.

 As was shown
recently, spontaneous vacuum magnetization appears in nonAbelian gauge
theories at high temperature. This was found by analytic methods
in   \cite{Starinets:1994vi}-\cite{Skalozub:1999bf} and was
confirmed by means of lattice simulations in \cite{Demchik:2008zz}. The
basic idea rests on the known fact that spontaneous vacuum magnetization is the
consequence of the  spectrum of a color charged gluon,
\begin{equation} \label{spectrum} p^2_{0} = p^2_{||} + (2 n + 1) g
B\qquad(n = - 1, 0, 1,... ), \end{equation}
in a homogeneous magnetic background, $B$, described  by the potential
\be\label{potentialB} A_\mu^a = B x_2 ~\delta_{\mu 3} \delta^{a 3},
\ee
where $a$ is the weak isotopic index,
and $p_{||}$ a momentum component along the field direction.
 Here, a tachyon  mode
 is present in the ground state ($n=-1$).  In fact, one observes that $p_0^2<0$, resulting
from the interaction of the magnetic moment of the spin-1 charged particles
with the magnetic field. This phenomenon was first discovered by
Savvidy \cite{Savvidy:1977as} at zero temperature, $T=0$, and got
known as the Savvidy vacuum. However, at zero and low temperature,
this state is not stable. It decays under emission of gluons until
the magnetic field $B$ disappears. This picture changes with
increasing temperature, where a stabilization sets in. The
stabilization is due to vacuum polarization and depends on two
dynamical parameters appearing for $T \not = 0$. These is a magnetic  mass of the color
charged gluon, $m_{magn.},$  and an $A_0$-condensate, which is proportional to the
Polyakov loop \cite{Ebert:1996tj}. This field configuration is stable,
its energy being below the perturbative one, and its minimum is
reached for the field being of order $g B \sim g^4 T^2/\log T$. This
phenomenon  is common to different $SU(N)$ gauge fields, which can
be used to extend the
standard $SU(2) \times U(1))_{ew} \times SU(3)_c$ model of
elementary particles.

An important property of such temperature-dependent magnetic
fields is the vanishing of their magnetic mass, $m_\mathrm{magn.}
= 0$. This was found both in one-loop analytic calculations
\cite{Bordag:2006pr} and in lattice simulations \cite{Antropov:2010}.
The mass parameter describes the inverse spatial scales of the
transverse field components, similarly to the Debye mass $m_D$,
related to the inverse space scale for the electric (Coulomb)
component. The absence of the screening mass means that the
spontaneously generated Abelian chromomagnetic fields are long
range at high temperature, as is common for the $U(1)$ magnetic
field. Hence, it is reasonable to believe that, at each stage
of the evolution of the hot Universe, spontaneously created,
strong, long-range magnetic fields of different types have been
present. Since they are constant fields, their scales are coinciding
with the horizon scale at each particular temperature. These fields
may have critically influenced various processes and phase transitions.

The  idea of the present investigation is to relate the generation of intergalactic
magnetic fields  with the reheating epoch which immediately followed the inflationary stage
of the universe evolution. At this epoch the temperature was the same in
all volume of the expanded space (as it follows from the spectra of relic photons),  and therefore, the field strength was also the same. Moreover,
since both inflation and the vacuum polarization are causal
processes, the coherence of the field
at very large scales can be ensured. On the other hand, the creation of strong temperature dependent fields at this stage completely
washed out the remnants of the fields generated at the inflation epoch.

The   dependence  on the temperature  of these  fields differs
from that of the usual  $U(1)$ magnetic fields.  Recall  that, in
the latter case, as is commonly assumed the magnetic (hypermagnetic, in fact) field,
created by some specific mechanism, is implemented in a hot plasma
and evolves according to the law $B \sim T^2$, which is a
consequence of magnetic flux conservation (see, for instance,
\cite{Kunze}). This scenario is actually considered, in magnetic hydrodynamics,
when the evolution
of large scale magnetic fields is investigated. The temperature dependence of a field
of this type is ordinarily related  with the high conductivity of the plasma. Thus, it is just a consequence of the classical theory of matter.  However, the classical theory is not always
sufficient for  describing the restored phase in models with spontaneous symmetry breaking. In this phase, quantum  effects are essential.
One of them is vacuum polarization, which results in spontaneous vacuum
magnetization. Just this phenomenon   regulates the dependence of the field strength on the temperature, and  the magnetic flux is {\it not} conserved. Instead,
 a specific  flux value is generated at each temperature.  This
fact has to be taken into consideration when the cooling pattern
of the hot  plasma is investigated at temperatures over that of the electroweak phase transition (EWPT), $T \geq T_{ew} \sim 100 $ GeV. This also concerns
the $SU(2)_{ew}$ component of the electromagnetic field. The classical theory is not able to  account for it consistently. Thus, classical magnetic hydrodynamics could only start to work after the vacuum magnetization has stopped. As we will show below, this does happen after the EWPT.

It is worth to note that one of the difficult problems of magnetogenesis is to relate the field strengths generated
via some given mechanism in the early Universe with those of the present-day fields, what depends on numerous factors and is very model dependent.
In magnetic hydrodynamics, the magnetic field evolution  was investigated in some detail
 for stochastic  fields presumedly  generated at either the electroweak or the confinement phase transitions (see \cite{BJ,KTR}, and references therein). It was shown that their fate is strongly dependent on the scales and initial conditions for these fields.
In particular, magnetic flux conservation could not hold for some special conditions. We make a detailed comparison of these results with those coming from our approach, in what follows. An important point  is that, in contrast with commonly investigated stochastic fields, we have here the possibility to deal with the evolution of a special solution of the field equations. Thus, its characteristics and properties are known and can be duely taken into consideration.

In the frameworks of the standard model, we will here
estimate the strength of the magnetic  field at the temperature
of the electroweak $T_c^\mathrm{ew}$ phase transition, assuming
that this field was spontaneously generated by a mechanism as
described above. Although such phenomenon is nonperturbative, we
carry out an actual calculation in the framework of a consistent
effective potential (EP) accounting for the one-loop, $V^{(1)}$,
and the daisy (or ring), $V^\mathrm{ring}$, diagram contributions.
In Sect.~2 we qualitatively describe, in quite more detail, the
main aspects of the investigated phenomena. In Sect.~3 the EP
of an Abelian constant electromagnetic $B$ field at finite
temperature is obtained. It is used, in Sect.~4, to estimate the
magnetic field strength at the EWPT temperatures. The scales of
the magnetic fields at high temperatures, and their
relation with the ones observed in intergalactic space, are
investigated in Sect.~5. A discussion of the results obtained,
some conclusions, and prospects for further work are provided
in the last section.

\section{Qualitative considerations}
In this section we describe, in a qualitative manner, the main
aspects of the phenomena at issue. They are all consequences
of  asymptotic freedom and spontaneous symmetry breaking  at
finite temperature.  Our main assumption is that the intergalactic
magnetic field has been spontaneously created at high temperature.
We believe, this to be a fairly reasonable idea because, physically,
magnetization is the consequence of a large magnetic moment for
charged non-Abelian gauge fields (recall the gyromagnetic ratio
$\gamma = 2$ for $W$-bosons). This property results in the
asymptotic freedom of the model in external fields.  We discuss
the procedure to relate the present value of the intergalactic
magnetic field with the one generated in the restored phase.

First, we note that, in non-Abelian gauge theories, magnetic flux
conservation does not hold at high temperatures, $T \geq
T_\mathrm{ew}$, where $T_\mathrm{ew}$ is the temperature of EWPT.
This is due to spontaneous  vacuum magnetization, which depends on
the temperature. The vacuum  acts as a specific source  generating
classical fields. Second, the  magnetization is strongly dependent
on the scalar field condensate present in the vacuum  at low
temperature. This point was investigated at zero temperature by
Goroku \cite{Goroku}.  For  finite temperature, this is
considered in the present paper for the first time. The observation is that, in both
cases, the spontaneous vacuum magnetization takes place for small values of the
scalar field, $\phi \not = 0$, only. For the values of $\phi$
corresponding to any first order phase transition  it  does not
happen. This means that, after the EWPT, the vacuum polarization
ceases to generate   magnetic fields, and magnetic flux
conservation holds.  As a result,  the familiar dependence on the
temperature, $B \sim T^2$, is restored (for the fields
spontaneously generated before the transition). However, that may be not the
case, in general, for stochastic magnetic fields generated just at
this transition \cite{BJ}. One has to distinguish these types of
fields, which have different nature and origin.

Another aspect of the problem is the composite structure of the
electromagnetic field $A_{\mu}$.  The potentials read
\begin{eqnarray} \label{AZ}
A_\mu &=& \frac{1}{\sqrt{g^2 + g'^2}} \left( g' A^3_\mu + g b_\mu
\right), \nonumber\\ Z_\mu &=& \frac{1}{\sqrt{g^2 + g'^2}} \left(
g A^3_\mu - g' b_\mu \right),
\end{eqnarray}
where $Z_\mu $ is the $Z$-boson potential, $A^3_\mu$, $b_\mu$ are
the Yang-Mills gauge field  third projection in  the weak isospin
space  and the potential of the hypercharge gauge fields, and $g$
and $g'$ are $SU(2)$ and $U(1)_Y$ couplings, correspondingly.
After the electroweak phase transition, the $Z$-boson acquires a
mass and the field is screened. Since the hypermagnetic field is
not spontaneously generated, only the component $A_\mu = g'
A^3_\mu/{\sqrt{g^2 + g'^2}} = \sin \theta_\mathrm{W}  A^3_\mu$ is
present at high temperature. Here $\theta_\mathrm{W} $ is the
Weinberg angle, $\tan \theta_\mathrm{W} = {g'}/{g}$. This is the
only component responsible for the intergalactic magnetic field at
low temperature.
In the restored phase, $b_\mu = 0$, the complete weak-isospin chromomagnetic field $A^{(3)}_\mu$ is unscreened. This is because  the magnetic mass of this field is zero \cite{Antropov:2010}. Thus, the field is of long range and this  provides  the coherence length $\lambda_B(T)$ to be sufficiently  large. In fact, because the field is a constant, it has to cover all the horizon scale at the given temperature, $\lambda_B(T) \sim R_{H(T)}$.
 This property is very important for our scenario of intergalactic magnetic field generation. Its cosmological consequences will be discussed below.
In particular, magnetic fields of different types (color $SU(3)$, and others) can be spontaneously generated at high temperatures.

 Here, we continue with the description of the general field behavior related with the EWPT.
In the restored phase, a scalar field condensate $\phi  = 0$, and  the constituent of the weak isospin field corresponding to the magnetic one,  is  given by the expression
\be \label{fieldT} B(T) = \sin \theta_w (T) B^{(3)}(T),\ee
where $B^{(3)}(T)$ is the  strength of the field generated
spontaneously.  After the phase  transition, the scalar condensate is
$\phi \not = 0$, and the field is partially screened.

To relate the present value of the intergalactic magnetic field
with  the field which existed before the EWPT, we take into
consideration  that, after the phase transition, spontaneous
vacuum magnetization does not take place. This  property is
derived in the next sections. Therefore, for the EWPT
temperature $T_\mathrm{ew}$, we can write
\be \label{relation} \frac{B(T_{ew})}{B_0} = \frac{T^2_{ew}}{T^2_0} = \frac{\sin \theta_w (T_{ew}) B^{(3)}(T_{ew})}{B_0}. \ee
Here, $B_0$ is the present value of the intergalactic  magnetic
field strength $B_0 \sim 10^{- 15} G$.  The left-hand-side relates
the value   $B(T_{ew})$ with  $B_0$. The right-hand-side gives a
possibility to express  the weak isospin magnetic field in the
restored phase  through $B_0$, knowing the temperature dependence
of the Weinberg angle $\theta_w(T)$. This relation  contains an
arbitrary temperature normalization parameter $\tau$. It can be
fixed for a given temperature and $B_0$.  After that,  the  field
strength values at various temperatures can be calculated. In
particular, the total weak isospin field strength is given by the
sum $ \cos \theta_w (T_{ew}) B^{(3)}(T_{ew}) + B(T_{ew})$.

Note also that  the relation \Ref{relation} is a consequence of
the  assumption that, for the field spontaneously generated at high
temperature, the magnetic flux conservation
holds, in the early Universe,  after the EWPT. This means that the field is ``frozen" in
the plasma at large scales, and the magnetic turbulence processes do
not affect that behavior. Although this is a most simple
assumption, it requires a detailed discussion encompassing the results
obtained recently in magnetic hydrodynamics (see Refs.~\cite{BJ,KTR},
and references therein). Details on the
magneto-hydrodynamical processes occurring in the early universe can be found in
numerous publications (see, for references, the review papers \cite{Giovannini,Kunze}).
Here, we will just mention the main points which are important for our
future considerations. The ``frozen in" conditions are always realized for   magnetic
fields having the  scales larger than  the largest turbulence eddies. After a
free decay stage of the magnetized plasma evolution,   the
 field can be considered to be non-affected by turbulence \cite{KTR}.
  In connection with these results,
it is clear that the magnetic fields generated at the EWPT are not
sufficient, as such, to produce long-range correlated fields, and
some additional  processes must be included. This is because, even
a field having the scale of the Hubble radius at temperature
$T_{ew} \sim 100$ GeV, is in fact correlated at the comoving scale $l_0
\sim 10^{- 4}$ Mpc \cite{KTR}.

But, in the case of a temperature-dependent vacuum polarization, other
possibilities exist. To obtain a   large-scale correlated
intergalactic magnetic field one can take into account that the
temperature in the Universe is the same, to high accuracy, and of the order $\delta
T \sim 10^{- 4}$. This is ensured by the reheating stage coming
just after  inflation. The temperatures at this stage are estimated
to be of the order $10^{16} - 10^{12}$ GeV depending on the specific
inflation model \cite{GorbunovRubakov}. Hence, with high accuracy,
the magnetic field strengths  $B(T)$ generated due to vacuum
polarization are also the same in all regions of the universe.
Since they are produced by a causal process, as inflation is, a
correlation of the fields at very large scale will result. In this
way, coherent magnetic fields on huge scales could be
generated. Such kind of fields have much larger scales than any
turbulence eddy, therefore, they will not be influenced by
turbulent processes. Thus, it is reasonable to assume that these
magnetic fields were frozen into plasma after the EWPT.

An important aspect of the present  scenario is that the knowledge of the particular theory, still unknown today, which may prove in future experiments to be the right extension to the standard model, is not that very important for estimating the field strength $B$ at
temperatures close to $T_{ew}$. This is so because the new gauge fields of the
   extended model will remain screened, at the high temperatures corresponding to the
   spontaneous symmetry breaking of some of the basic symmetries.  At very high temperatures, when these symmetries will be restored, the corresponding magnetic fields do will exist. In this way the value of the field strength at the Planck era can be  estimated (see Ref.~\cite{Pollock}, for comparison).
\section{Effective potential at high temperature}
As we noted above, the spontaneous vacuum magnetization and the
absence of the magnetic mass for the Abelian magnetic fields are
nonperturbative effects to be determined, in particular, in
lattice simulations \cite{Demchik:2008zz,Antropov:2010}. The main
conclusions of these investigations are that a stable magnetized
vacuum does exist at high temperature and that the magnetic mass
of the created field is zero. Concerning the actual value of the
field strength, is is close to the one calculated within the
consistent effective potential which takes into account one-loop
plus daisy diagrams. Thus, in the present investigation we
restrict ourselves to such approximation. This is mainly with the
purpose to be able to use analytic calculations in order to
properly interpret the results.

The complete EP for the standard model is given in the review
\cite{Demchik:1999}. In the present investigation we are
interested in two limits,
\begin{enumerate}
\item weak magnetic field and large scalar field condensate,
$h = eB/M_W^2 < \phi^2$, $\phi = \phi_c/\phi_0$, $\beta = 1/T$;
\item the case of the restored symmetry, $\phi = 0$, $g B \not = 0$,
$T \not = 0$.
\end{enumerate}
For the first case, we show the absence of spontaneous vacuum
magnetization at finite temperature. For the second, we
estimate the field strength at high temperature. Here $M_W $ is
the $W$-boson mass at zero temperature, $\phi_c $  a scalar
field condensate, and $\phi_0$ its value at  zero temperature.

To demonstrate the first property we consider the one-loop
contribution of  $W$-bosons (see also Eq.~(27) of
Ref.~\cite{Skalozub:1996ax}),
\begin{eqnarray} \label{L2t}
V^{(1)}_W(T,h,\phi) &=& \frac{h}{\pi^2 \beta^2} \sum\limits_{n =
1}^{\infty} \left[ \frac{\beta\sqrt{\phi^2 - h}}{n} K_1\left(n
\beta \sqrt{\phi^2 - h}\right)\right. \nonumber\\ &&\left.-
\frac{\beta\sqrt{\phi^2 + h}}{n} K_1\left(n \beta \sqrt{\phi^2 +
h}\right)\right].
\end{eqnarray}
Here $n$ labels discrete energy values and $K_1(z)$ is the MacDonald function.

The main goal of our investigation is the restored phase of the
standard model. So, we obtain the high temperature contribution of
the complete effective potential relevant for this case using the
results in Ref.~\cite{Demchik:1999}. First,  we write down the
one-loop $W$-boson contribution as the sum of the pure Yang-Mills
weak-isospin part $(\tilde{B}\equiv B^{(3)}$),
\begin{eqnarray} \label{VW}
V^{(1)}_W(\tilde{B}, T) &=& \frac{\tilde{B}^2}{2} +  \frac{11}{48}
\frac{g^2}{\pi^2} \tilde{B}^2 \log \frac{T^2}{\tau^2} -
\frac{1}{3} \frac{( g \tilde{B})^{3/2} T }{\pi} \nonumber\\  &-& i
\frac{( g \tilde{B})^{3/2} T }{2 \pi} + O (g^2 \tilde{B}^2),
\end{eqnarray}
where $\tau$ is a temperature normalization point, and the charged
scalars  \cite{Skalozub:1996ax},
\begin{equation} \label{Vscalar}
V^{(1)}_\mathrm{sc}(\tilde{B}, T) =  - \frac{1}{96}
\frac{g^2}{\pi^2} \tilde{B}^2 \log \frac{T^2}{\tau^2} +
\frac{1}{12} \frac{( g \tilde{B})^{3/2} T }{\pi} + O (g^2
\tilde{B}^2),
\end{equation}
describing the contribution of longitudinal vector components. The
first term in Eq.~\Ref{VW}  is the tree-level energy of the field.
This representation is convenient for the case of extended models
including other gauge and  scalar fields. Depending on the
specific case,  one can take into consideration  the parts
\Ref{VW} and \Ref{Vscalar}, correspondingly. In the standard model,
the contribution of Eq.~\Ref{Vscalar} has to be taken with a
factor 2, due to the two charged scalar fields entering the scalar
doublet of the model. In the case of the Two-Higgs-Doublet
standard model, this factor must be 4, etc. The imaginary part is
generated because of the unstable mode in the spectrum
\Ref{spectrum}. It is canceled by the term appearing in the
contribution of the daisy diagrams for the unstable mode
\cite{Skalozub:1999bf},
\be \label{ringdaisy} V_{unstable} = \frac{g \tilde{B} T}{2 \pi} [\Pi(\tilde{B}, T, n = - 1) - g \tilde{B} ]^{1/2} + i \frac{(g \tilde{B})^{3/2} T}{2 \pi}. \ee
Here $\Pi(\tilde{B}, T, n = - 1)$  is the mean value  for the
charged gluon polarization tensor taken in the ground state $ n =
- 1$ of the spectrum \Ref{spectrum}. If this value is sufficiently
large, spectrum stabilization due to radiation correction takes
place. This possibility formally follows from the temperature and
field dependence of the polarization tensor in the high
temperature limit $T \to \infty $ \cite{Bordag08}: $\Pi(\tilde{B},
T, n = - 1)= c~ g^2 T \sqrt{g \tilde{B}} $, where $c > 0$ is a
constant which must be calculated explicitly. At high temperature
the first term can be larger then $g \tilde{B}$.
 From Eqs.~\Ref{VW} and \Ref{ringdaisy} it follows that the imaginary part is
 canceled. Hence, we see  that having accounted for rings leads
 to vacuum stabilization even if $ \Pi(\tilde{B}, T, n = - 1)$
 is smaller then $g \tilde{B}$. Really, in the latter case, the imaginary
 part will be smaller than in Eq.~\Ref{VW}.
 The high temperature limit of the fermion contribution looks as
 \be \label{fermionEP} V_{fermion} = - \frac{\alpha}{\pi} \sum\limits_{f} \frac{1}{6} q^2_f \tilde{B}^2 \log\frac{ T}{\tau} , \ee
 where the sum is extended to all leptons and quarks, and $q_{f}$
is the fermion electric charge in positron units.    Hence, it
follows that in the restored phase all the fermions give the same
contribution.

 Let us now present the EP for ring diagrams  describing the long
range correlation corrections at finite temperature
\cite{Carrington:1992,Demchik:2003},
\begin{eqnarray} \label{Vring}
V_\mathrm{ring} &=& \frac{1}{24 \beta^2} \Pi_{00}(0)  - \frac{1}{12 \pi \beta}
\,\mathrm{Tr} [\Pi_{00}(0)]^{3/2}\nonumber \\
&+& \frac{(\Pi_{00}(0))^2}{32 \pi^2} \left[\log\frac{4 \pi}{\beta
(\Pi_{00}(0))^{1/2}} + \frac{3}{4} - \gamma \right],
\end{eqnarray}
where the trace means summation over all the contributing states,
$\Pi_{00} = \Pi_{\phi}(k = 0, T, B)$ for the Higgs particle;
$m_D^2 = \Pi_{00} = \Pi_{00}(k = 0, T, B)$ are the zero-zero
components of the polarization functions of gauge fields in the
magnetic field taken at zero momenta, called the Debye mass
squared, and $\gamma$ is Euler's gamma. These terms are of order $\sim
g^3 (\lambda^{3/2})$ in the coupling constants. The detailed
calculation of these functions is given in
Ref.~\cite{Demchik:1999}. We give the results for completeness,
\bea \label{Piscalar} \Pi_{\phi}(0) &=& \frac{1}{24 \beta^2} \bigl( 6 \lambda + \frac{6 e^2}{\sin^2 (2\theta_w)} + \frac{3 e^3}{\sin^2 \theta_w} \bigr)\\ \nn &+& \frac{2 \alpha}{\pi} \sum\limits_f \bigl[ \frac{\pi^2 K_f}{3 \beta^2} - |q_f B| K_f \bigr] \\ \nn &+& \frac{(e B)^{1/2}}{8 \pi \sin^2 \theta_w \beta} e^2 (3 \sqrt{2} \zeta(-\frac{1}{2}, \frac{1}{2})). \eea
Here $K_f = \frac{m_f^2}{M_w^2} = \frac{G^2_{Yukawa}}{g^2}$ and $ \lambda $ is the scalar field coupling.
The terms \mbox{$\sim
T^2$} yield standard contributions  to the temperature mass squared
coming from the boson and fermion sectors. The $B$-dependent terms
are negative (note the value of $3 \sqrt{2} \zeta(-\frac{1}{2},
\frac{1}{2}) = - 0.39$). They  decrease the value of the screening
mass at high temperature. The Debye masses squared for the
photons, $Z$-bosons, and neutral current contributions are,
respectively,
\begin{eqnarray} \label{mAZ}
&& m^2_{D, \gamma} = g^2 \sin^2\theta_\mathrm{W}
\left[\frac{1}{3 \beta^2} + O(e B \beta^2)\right], \nonumber\\
&& m^2_{D, Z} = g^2 \left(\tan^2\theta_\mathrm{W} + \frac{1}{4
\cos^2\theta_\mathrm{W}} \right)
\left[\frac{1}{3 \beta^2} + O(e B \beta^2)\right], \nonumber\\
&& m^2_{D,\mathrm{neutr.}} = \frac{g^2}{ 8
\beta^2\cos^2\theta_\mathrm{W} } \left(1 + 4 \sin^4
\theta_\mathrm{W}\right) + O(e B \beta^2).\
\end{eqnarray}
As one can see, the dependence on $B$ appears at order $O(T^{-2})$.

The $W$-boson contribution to the Debye mass of the photons is
\begin{equation} \label{mW}
m^2_{D, W} = 3 g^2 \sin^2\theta_\mathrm{W}
\left(\frac{1}{3 \beta^2} - \frac{(g \sin \theta_\mathrm{W}
B)^{1/2}}{2 \pi \beta}\right).
\end{equation}
An interesting feature of this expression is the negative sign of the
next-to-leading terms which dependen on the field strength. Finally, we
give the contribution of the high temperature part in
Eq.~\Ref{ringdaisy} $\Pi(\tilde{B}, T, n = - 1)$
\cite{Demchik:1999},
\begin{equation} \label{Piunstable}
\Pi(\tilde{B}, T, n = - 1) = \alpha \frac{(g \sin
\theta_\mathrm{W} B)^{1/2}}{\beta} \left( 12.33 + 4i \right).
\end{equation}
This expression was calculated from the one-loop $W$-boson
polarization tensor in the external field at high temperature. It
contains the imaginary part which comes from the unstable mode in
the spectrum \Ref{spectrum}. Its value is small as compared to the
real one. It  is of the order of the usual damping constants in
plasma at high temperature. It will be thus ignored in actual
calculations in what follows. In fact, this part must be
calculated in a more consistent scheme which starts with a
regularized stable spectrum. On the other hand, as we noted above,
the stability problem is a non-perturbative one. Stabilization
can be realized not only through radiation corrections but also by
some other mechanisms. For example, due to $A_0$ condensation
\cite{Starinets:1994vi} at high temperature. A
stable vacuum state was observed in lattice simulations
\cite{Demchik:2008zz}, therefore, we believe that this problem has
a positive solution. Summing up, we have all necessary ingredients to
investigate the problem of interest.

\section{Magnetic field strength at $T_{ew}$ }
Let us now show that the spontaneous vacuum magnetization does not
take place at finite temperature and for non-small values of the
scalar field condensate $\phi \not = 0$. To this end we notice that
the magnetization is produced by the gauge field contribution,
given in Eq.~\Ref{L2t}. So, we consider the limit of ${g B}/{T^2}
\ll 1$ and $\phi^2 > h$. For this case we use the asymptotic
expansion of $K_1(z)$,
\begin{equation} \label{K1asympt}
K_1(z) \sim \sqrt{\frac{\pi}{2 z}} e^{- z} \left( 1 + \frac{3}{8
z} - \frac{15}{128 z^2} + \ldots \right),
\end{equation}
where $ z = n \beta (\phi^2 \pm h)^{1/2}$. Now, we investigate the
limit of $\beta \to \infty$, ${T}/{\phi} \ll 1$, where the
leading contribution  is given by the first term of the
temperature sum in Eq.~\Ref{L2t}. We can also substitute $(\phi^2
\pm h)^{1/2} = \phi ( 1 \pm \frac{ h}{2 \phi^2})$. In this
approximation, the sum of the tree level energy and \Ref{L2t}
reads
\begin{equation} \label{L2tasympt}
{V} = \frac{h^2}{2} - \frac{h^2}{\pi^{3/2}}
\frac{T^{1/2}}{\phi^{1/2}} \left( 1 - \frac{T}{2 \phi} \right)
e^{- {\phi}/{T}}.
\end{equation}
The second term is exponentially small and the stationary equation
${\partial{V}}/{\partial{h}} = 0$ has the trivial solution $h =
0$. This estimate can be easily verified in a numerical calculation
of the total effective potential. Hence, we conclude that after
symmetry breaking the spontaneous vacuum magnetization does not
take place, as was the case at zero temperature \cite{Goroku}.

To  estimate the magnetic field strength in the restored phase at the EWPT temperature the total EP deduced in the previous section must be used and the parameters entering Eq.~\Ref{relation} need to be calculated. This can be best done numerically. To explain the procedure, we consider here a part of this potential accounting for the one-loop $W$-boson contributions. The high temperature expansion for the EP coming from charged vector fields is given in Eq.~\Ref{VW}.  Assuming stability of the vacuum state, we calculate the value of the chromomagnetic weak isospin field spontaneously generated at high temperature from Eqs.~\Ref{VW} and \Ref{Vscalar}:
\be \label{fieldT1} \tilde{B}(T) = \frac{1}{16} \frac{g^3}{\pi^2} \frac{T^2}{(1 + \frac{5}{12} \frac{g^2}{ \pi^2} log \frac{T}{\tau})^2 }.\ee
This expression (and the complete one accounting for all contributions) gives the field strength  at any temperature, $T \geq T_{ew}$. These formulas can be obtained for different types of particles. Before reducing to a specific value for it,
we  describe  how to connect this expression with the intergalactic magnetic field $B_0$. We  first  relate the expression \Ref{fieldT1} with an electromagnetic field after symmetry breaking, and then  take into account the scales of the fields.

Let us introduce the standard parameters and definitions,
$\alpha_\mathrm{w}={g^2}/{(4 \pi)}$, $\alpha =  \alpha_\mathrm{w}
\sin \theta_\mathrm{W}^2$, $\alpha_Y={(g')^2}/{(4 \pi)}$ and
$\tan^2 \theta_\mathrm{W}(T) = { \alpha_Y(T)}/{
\alpha_\mathrm{w}(T)}$, where $\alpha$ is the fine structure
constant. To find the temperature dependence of the Weinberg
angle, the behavior of the hypercharge coupling $g'$  on the
temperature has to be computed. From Eq.~\Ref{Vscalar} it follows
that this behavior is nontrivial. The logarithmic
temperature-dependent term is negative. But, as is well known, in
the asymptotically free models this sign must be changed into a
positive value, due to the contributions of other fields. This
particular value is model dependent. We will not calculate it in
the present paper. Instead, for a rough estimate, we substitute
the zero-temperature value: $\sin^2 \theta_\mathrm{W}(T) = \sin^2
\theta_\mathrm{W}(0)= 0.23$.

For the given temperature of the EWPT, $T_\mathrm{ew}$, the
magnetic field is
\begin{equation} \label{B3T}
B(T_\mathrm{ew}) = B_0 \frac{T^2_\mathrm{ew}}{T^2_0} = \sin
\theta_\mathrm{W} (T_\mathrm{ew})
\tilde{B}(T_\mathrm{ew}).
\end{equation}
Assuming $T_\mathrm{ew} = 100 GeV =    10^{11} eV$ and $T_0 = 2.7
K = 2.3267 \cdot 10^{- 4} eV$, we obtain
\begin{equation} \label{Bew}
B(T_\mathrm{ew}) \sim 1.85 ~10^{14} G.
\end{equation}
This value can be considered as an estimate of the magnetic field strength at the EWPT.
Hence, for the value of $X =  \log \frac{T_{ew}}{\tau}$, we have the equation
\be \label{Xew} B_0 =  \frac{1}{2}  \frac{\alpha^{3/2}  }{\pi^{1/2} \sin^2 \theta_w } \frac{T^2_0}{(1 + \frac{5 \alpha}{3 \pi \sin^2 \theta_w }  X)^{2}}. \ee
Since all the values are known, $\log \tau $ can be computed.
After that, the field strengths at different higher temperatures
can be found. In fact, the main point in obtaining these results is the assumption of magnetic flux conservation, as is frozen in a plasma. Information of a particular model is implemented in the factor $\sin \theta_w (T_{ew})$ in Eq.~\Ref{B3T}.
Needles to say, our estimate is a rough one, because we have ignored
the temperature dependence of the Weinberg angle. To guess the
value of the parameter $\tau$ we take  the field strength $B_0
\sim 10^{- 9} G$, usually used in cosmology (see, e.g.,
\cite{Pollock}). In this case, from Eq.~\Ref{Xew} we obtain $\tau
\sim 300 eV$. For the present-day value $B_0 \sim 10^{- 15}G$ this
parameter is much smaller.

To take into account the fermion contribution
Eq.~\Ref{fermionEP}, we have to substitute the expression
$\frac{5}{12} \frac{g^2}{ \pi^2} log \frac{T}{\tau}$ in
Eq.~\Ref{fieldT1} and also in  \Ref{Xew}, with the value
\be
\label{plusferm} (\frac{5}{3} -  \sum\limits_{f} \frac{1}{6} q^2_f   )\frac{\alpha_s}{ \pi} log \frac{T}{\tau}.
\ee
In the above estimate, we have accounted for the  one-loop
part of the EP of order $\sim g^2$ in the coupling constant. The
ring diagrams have order $\sim g^3$ and give a small numeric
correction to  this result.
As was mentioned before, had we taken into account all the
terms listed in the previous section, the results would have not
changed essentially.

Let us compare now the value of the field strength \Ref{Bew} with the
one calculated directly from the EP for the standard model in
Ref.~\cite{Demchik:2002}.  From Fig.~1 and Tab.~1 of that paper, we find
\be
\label{BewSM} B^{SM}(T_{ew}) \sim  10^{20} G,
\ee
what is much larger than the value \Ref{Bew} and just corresponds to the value of the present comoving
field strength $B_0 \sim 10^{- 9} G$. Note that this value was used in numerous investigations of magnetic fields
in the early universe, before the recent discoveries \cite{Ando}-\cite{Ando2}.
Let us stress again that the field strength at higher temperatures will depend on the
particular model extending the standard one. Spontaneous vacuum
magnetization in the minimal supersymmetric standard model has
been investigated in Ref.~\cite{Demchik:2003}, and the field strength
generated in this model is smaller as compared to the situation here
considered. Also,  Pollock   \cite{Pollock} has investigated this
problem   for the case of the Planck era, where magnetic fields
of the order $B \sim 10^{52} G$ have been estimated.
\section{Magnetic field scale}
We now discuss in brief the scale of the field generated in the restored phase. This is a key point in  relating expressions like Eqs.~\Ref{fieldT1} or \Ref{BewSM} with $B_0$. In our consideration, the ``frozen in" condition was used. Therefore, we are going  to discuss its applicability in more detail. Note first that, if one assumes that after the EWPT   the constant  field  $B(T_{ew})$  was frozen in the plasma at the Hubble scale, $R_H(T_{ew})$, then its  comoving coherence scale at present will be $\la_B(T_0) = 6 \cdot 10^{-4}$ pc \cite{KTR}. This is much smaller than is needed.

We propose two, in fact related,  ways to overcome such difficulty. The first is to take into consideration the reheating stage of the universe evolution. According to the concepts of modern cosmology \cite{GorbunovRubakov},  this stage has existed just after inflation and  is related with the latter causal stage.    Just due to causality, the temperature in the universe after this stage is the same, in all  domains of space, which could even be uncorrelated in later moments of time.
Hence, at a given high temperature, $T$, the   magnetic field generated due to vacuum polarization has the same strength $B(T)$ everywhere in the universe. Formally, they could have different  directions, in either external or internal spaces, although this point requires additional consideration.  Different kind of chromomagnetic fields, of the type as in Eq.~\Ref{potentialB}, can be  spontaneously generated.  Their nature depends on the particular model considered and is therefore unknown, as of now. But this is not essential for our consideration, here. The magnetic fields coherent on huge scales are expected to have been present in the early Universe. The origin of this coherence is ensured by the properties of the solution to the field equations (Eq.~\Ref{potentialB})  and  by causality at the inflationary epoch. The scales of the coherent field domains could be estimated on grounds of the gauge invariance. This idea, due to Feynman, was put in force in gluodynamics  with the goal to determine  possible magnetic vacuum structures \cite{Feynman}. Namely, to find a gauge invariant vacuum, on the basis of gauge non-invariant solutions (such as Eq.~\Ref{potentialB}), one can consider a domain  structure ensuring gauge invariance when a corresponding boundary is going around. This point requires further investigation.

 Most of the fields generated in the early Universe decouple and are screened at some energy (temperature) scales,  when the corresponding scalar condensates have broken the background symmetries. So, the only unbroken symmetry at the EWPT remains the $SU(2)_{ew}\times U(1)$ one. After the EWPT, when spontaneous magnetization stops, this field cannot  be included in turbulent processes generated by the transition. This is  because the scale of the  field, Eq.~\Ref{potentialB}, is already much larger than any largest  eddy of turbulence. As is usually believed, the size of a typical eddy is estimated as the inverse mass of the particles appearing after the transition \cite{Vachaspati}. Thus, the field  evolves in accordance with  the metric expansion and is implemented in a hot plasma, thus fulfilling the magnetic flux conservation law.  And it finally results in the present day  intergalactic magnetic fields which could be correlated on $\sim  1$  Mpc scales. Note that an essential information on  the processes that take place after the EWPT, obtained in the framework of magneto-hydrodynamics, is given in Refs.~\cite{KTR} and \cite{Barrow}. We will discuss the facts which  have relevance to our problem in the last section.

Another possible scenario is based on the stochastic processes considered already by Hogan \cite{Hogan83} in connection with the magnetic fields generated at first-order EWPT. A possible mechanism of field generation in that case was proposed by Vachaspati \cite{Vachaspati}. In the former paper, it was pointed out that magnetic fields correlated on large scales can be produced not only through causal processes but also by a stochastic random walk mechanism, if the magnetic lines generated in some domain of space ``forget" about their origin. The field strength developed on large scales by this process (due to ``straightening" of entangled magnetic fields)    can be estimated as $B_N \sim  B/\sqrt{N}$, where $N$ counts the number of domains, with the field $B$ of a given size, crossed by a magnetic line. The correlation length $\la_B$ in this case can be much larger than the $R_H(T)$. It can be estimated as $\la_B(T) \sim N R_H(T)$. In the paper \cite{Hogan83}, it has been also noticed that this mechanism is not applicable to the early universe, the reason being because  magnetic lines do not penetrate freely though the plasma. This is really the case, if the properties of the plasma as such are taken into account. However, this is not the  case if  spontaneous vacuum magnetization occurs. Actually, at a given temperature, each uncorrelated domain of space having a Hubble radius $R_{H}(T)$ is filled up with a constant magnetic field $B(T)$, described by  the potential \Ref{potentialB}. Its orientation in both external and internal spaces is arbitrary. Hence, a stochastic behavior of the field lines and the appearance of  magnetic fields having large correlation lengths  $\la_B(T) \ge R_H(T)$ are expected. After the EWPT, these fields evolve as in the previous case. 

Note that, in both scenarios, all the fields generated at the inflation epoch are washed out by the vacuum polarization and leave no remnants at present. The reheating stage becomes more important. In closing this section, we also notice that the long range nature of the Abelian spontaneously generated magnetic fields is ensured by their zero magnetic mass (see \cite{Antropov:2010,Bordag:2006pr}), what renders these fields unscreened, as is the case for usual $U(1)$ fields.

\section{Discussion and conclusions}
We here summarize our main results. The key point in the problem under investigation  is the spontaneous vacuum magnetization process,  which eliminates the magnetic flux conservation principle at high temperature. Vacuum polarization is responsible for the value of the field strength $B(T)$ at each temperature and serves as its source. In fact, it converts heat into an ordered coherent state, which was not taken into consideration in previous studies of the early Universe. We have also shown here that, at finite temperature and after symmetry breaking, a scalar field condensate suppresses the magnetization.  Hence, it follows that the actual nature of the model extending the standard one is {\it not} that essential at sufficiently low temperatures, when  the decoupling of the heavy gauge fields has occurred.  From this one can conclude, in particular, that the vacuum polarization ``washes out" the relics of the magnetic fields generated at very high temperature or at the inflation stage.   These  statements are new and come as an interesting surprise, as compared with the standard notions based on the ubiquitous scenario with magnetic flux conservation.

The present value of the intergalactic magnetic field strength has been here related with fields at high temperatures in the restored phase. Because of the zero magnetic mass for Abelian magnetic fields (as discovered recently \cite{Bordag:2006pr,Antropov:2010}),  there is no problem for the generation of fields having  large coherence scales. In our estimates, we have assumed that, basically, the field  is  of the order of the horizon scale, $\la_B(T) \sim R_{H(T)}$.  This seems reasonable because, at a given temperature, the field $B(T) = const$,  generated due to vacuum polarization, occupies all space. Then, at the reheating stage (due to causality, present already at inflation), coherence on scales exceeding that of the horizon can be produced. Such large scale fields are not influenced by turbulent processes happening after the EWPT. They are frozen in the plasma and evolve according to the magnetic flux conservation law.  In this scenario, a large scale domain structure is also permissible, what requires additional consideration.

Knowing the particular properties of the extended model, it is possible to estimate the field strengths at any temperature. This can be done for different schemes of spontaneous symmetry breaking (restoration) by taking into account the fact that, after the decoupling of some massive gauge fields, the corresponding magnetic fields are screened. Thus, the higher the temperature, the larger the number of strong long-range magnetic fields of different types that will be generated in the early Universe.

As we have found above, the field strengths at the EWPT temperature, estimated with account to the present-day value of the intergalactic magnetic field strength, $\sim 10^{-15}$ G,   Eq.~\Ref{Bew}, or either directly from the vacuum magnetization in the  standard model, Eq.~\Ref{BewSM}, differ in six  orders of magnitude. This huge  deviation can be explained  by the different scales of the fields considered. Let us check this possibility by using the second of the scenarios proposed in the previous section, for large scale field generation. Making use of the usual relation between the scale factor and the temperature,
\be
\label{a-T} \frac{a(T_{ew})}{a(T_{0})} = \frac{T_{0}}{T_{ew}},
\ee
taken at the EWPT epoch, and the present-day parameters, $T_{ew}= 100 ~GeV = 10^{11} ~ eV , T_0 = 2.3267 \cdot 10^{- 4}$ eV. If one assumes that  $ ~\la_B(T_{}) \sim a(T_{})$, then from \Ref{a-T}, it follows that $\la_B(T_0) = 6 \cdot 10^{-4}$ ps (see, for instance, \cite{KTR}). On the other hand, if one takes $\la_B(T_0) = 1 $ Mpc, the value $\xi_B(T_{ew}) = 2.33 \cdot 10^{- 15}$ Mpc is obtained. At the same time, the horizon size is $a(T_{ew})= 1.27 \cdot 10^{-24} $ Mpc, thus, $\la_B(T_{ew}) >> a(T_{ew})$. Now, following an idea of Hogan \cite{Hogan83}, we relate the size of the correlated field with the random walk process. At $T_{ew}$, we have $\la_B(T_{ew}) = N a(T_{ew})$, hence, we get roughly $\sqrt{N} = 3 \cdot 10^4$, and for the field strength ``straightened" on the $N$-domain scale,  $B_N \sim \frac{B(T_{ew})}{\sqrt{N}}$ \cite{Hogan83}. Therefore, accounting for the field strength value calculated for the standard model, Eq.~\Ref{BewSM}, we obtain $B_{ls}(T_{ew}) \sim 3 \cdot 10^{15}$ G (the subscript in $B_{ls}$ means ``large scale"). This value is close the the value $B_{ls}(T_{ew}) \sim 2 \cdot 10^{14}$ G estimated in Eq.~\Ref{Bew}. The  remaining discrepancy can be explained in two ways. First, and obviously, as due to the roughness of our estimate. Second, and more radically, by the necessity of substituting the standard model with another one. The latter point will be discussed in more detail below.

Let us note a number of properties of the field under consideration, and compare them with the ones usually discussed in applications of magnetohydrodynamics to the early Universe.  Here, we follow  Refs.~\cite{KTR} and \cite{Barrow} which are close to our analysis. First, let us mention that the field energy density, $\rho_B = \frac{B^2}{2}$, is proportional to $g^6 T^4$, what is  much smaller than the radiation energy density, $\sim T^4$. Thus, the BBN condition (see \cite{KTR}), $\rho_B/\rho_{rad.} << 1$, is fulfilled. Second, as numerical simulations show \cite{KTR}, the turbulent process in the early Universe with magnetic field included is slower then in the laboratory.  Turbulence can include a large scale field  at the level of the largest eddies. For large scale fields, the free decay stage is important. At this epoch, which is strongly dependent on the initial conditions \cite{Barrow}, turbulence is significantly decreasing, and after this very brief stage the field is not affected by turbulence any more. It is just frozen in the plasma. As we have shown, the spontaneous vacuum magnetization is stopped when the first-order phase transition ends. Fields of this type cannot be influenced by turbulence and it is thus reasonable to believe that after the EWPT the field evolves  according to Eq.~\Ref{relation}. Note also that these fields are non-helical ones.

Our analysis has shown that, at the EWPT temperature, magnetic fields of the order $B(T_{ew}) \sim 10^{14} G$ did very likely exist. To estimate their field strengths at higher temperatures, one has to take into consideration a number of features proper to the standard model and its particular extension at play. First,  we note that quarks possess both electric and color charges. Therefore, there is a mixing between the color and usual magnetic fields owing to the quark loops. Second, there are   peculiarities related with the  particular content of the extended models. For example, in the Two-Higgs-Doublet standard model the contribution $\sim (g B)^{3/2} T$ in Eq.~\Ref{VW} is exactly canceled by the corresponding term in Eq.~\Ref{Vscalar}, because of the four charged scalar fields entering the model. They interact with gauge fields with the same coupling  constant. However, in this model the doublets interact differently with fermions. This changes the effective couplings of the doublets with the gauge fields and results in non-complete cancelations.  As a result, a  suppression of the spontaneously created magnetic field is expected in this model. In principle, one should be able to explain, in this way, the discrepancy in the field strengths as discussed above.
There can be other peculiarities which may influence the high temperature phase of the universe. They will require further investigation and thus we leave this issue for a future publication.
\vspace*{5mm}

\noindent{\bf Acknowledgements.}
The authors are grateful to Michael Bordag for numerous discussions and
a careful reading of the manuscript.
VS was supported by the European Science
Foundation CASIMIR Networking program. He also thanks the Group of
Theoretical Physics and Cosmology, at the Institute for Space
Science, UAB, Barcelona, for kind hospitality. EE's research was partly
carried out while on leave at the Department of Physics and Astronomy,
Dartmouth College, 6127 Wilder Laboratory, Hanover, NH 03755, USA.
This work has been also partly supported by MICINN (Spain),
projects FIS2006-02842 and FIS2010-15640, and Contract PR2011-0128,
by the CPAN Consolider Ingenio Project, and by
AGAUR (Generalitat de Ca\-ta\-lu\-nya), contract 2009SGR-994.


\end{document}